%% file: 0_Main.tex
\documentclass[submission]{eptcs}[10pt]
\providecommand{\event}{FTSCS 2012}

\usepackage{graphicx,amssymb,amstext,amsmath,amsthm,hyperref,indentfirst}
\usepackage{array,amsfonts, amsmath, amssymb}
\usepackage{xspace}
\usepackage[utf8]{inputenc}
\usepackage[T1]{fontenc}
\usepackage{ulem}
\usepackage{paralist}
\usepackage{caption}
\usepackage{subcaption}
\usepackage{multicol}
\usepackage{wrapfig}
\usepackage{algorithm}
\usepackage{algorithmicx}
\usepackage{algpseudocode}
\usepackage{tikz}

\hypersetup{
  pdfborder={0 0 0},
  colorlinks=true,
  linkcolor=blue,
  citecolor=blue,
  urlcolor=blue,
  pdftitle={
    Generating Property-Directed Potential Invariants By Backward Analysis
  }
}

\title{
  Generating Property-Directed Potential Invariants By Backward Analysis
}
\author{
  Adrien Champion
  \institute{Onera, The French Aerospace Lab\\Toulouse, France}
  \institute{Rockwell Collins France\\Blagnac, France}
  \email{adrien.champion@onera.fr}
  \and
  R\'emi Delmas
  \institute{Onera, The French Aerospace Lab\\Toulouse, France}
  \email{remi.delmas@onera.fr}
  \and
  Michael Dierkes
  \institute{Rockwell Collins France\\Blagnac, France}
  \email{mdierkes@rockwellcollins.com}
}
\def\titlerunning{Property-Directed Lemmas By Backward Analysis}
\def\authorrunning{A. Champion, R. Delmas, M. Dierkes}

\begin{document}
\normalem

\maketitle

\begin{abstract}
  This paper addresses the issue of lemma generation in a $k$-induction-based
  formal analysis of transition systems, in the linear real/integer arithmetic fragment.
  A backward analysis, powered by quantifier elimination, is used to output
  preimages of the negation of the proof objective, viewed as unauthorized states, or \emph{gray states}.
  Two heuristics are proposed to take advantage of this source of information.
  First, a thorough exploration of the possible partitionings of the gray state space
  discovers new relations between state variables, representing potential invariants.
  Second, an inexact exploration regroups and over-approximates disjoint areas
  of the gray state space, also to discover new relations between state variables.
  $k$-induction is used to isolate the invariants and check if they strengthen
  the proof objective.
  These heuristics can be used on the first preimage of the backward exploration, and each time a new one
  is output, refining the information on the gray states.
  In our context of critical avionics embedded systems,
  we show that our approach is able to outperform other academic or commercial
  tools on examples of interest in our application field.
  The method is introduced and motivated through two main examples, one of which was provided by
  Rockwell Collins, in a collaborative formal verification framework.
\end{abstract}

\section{Introduction}
\input{Introduction}

\section{Fault Tolerant Avionics Architectures}
\input{Systems}

\section{Related Work and Tools}
\input{Context}

\section{Notations}
\input{Notations}

\section{Proofs by Temporal Induction}
\input{kind}

\section{Approach Overview: Backward Exploration and Hull Computation}
\input{Approach}

\section{Generating Potential Lemmas Through Hull Computation}
\input{Hullqe}

\section{Applications}
\input{Applications}

\section{Framework and Implementation}
\input{Framework}

\section{Conclusion}
\input{Conclusion}

\bibliographystyle{eptcs}
\bibliography{Biblio}

\end{document}

%% file: Introduction.tex
\label{intro}
The recent DO-178C and its formal methods supplement DO-333 published by
RTCA\footnote{\href{http://www.rtca.org/}{http://www.rtca.org/}}
acknowledge the use of formal methods for the verification and
validation of safety critical flight control software and allow
their use in development processes. Successful
examples of industrial scale formal methods applications exist, such
as the verification by Astr\'ee \cite{BlanchetEtAl-PLDI03} of the run-time safety of the Airbus A380 flight control software C code.
However, the verification of general functional properties at model level, {\em i.e.}  on
Lustre~\cite{DBLP:conf/popl/CaspiPHP87} or MATLAB Simulink{\textcopyright} programs, from which the
embedded code is generated, still requires expert human intervention to succeed on
common avionics software design patterns,
preventing industrial designers from using formal verification on a larger scale.
Formal verification at model level is important, since it helps raising the confidence in the correctness of the
design at early stages of the development process.
Also, the formal properties and lemmas discovered at model level can be forwarded to the generated code, in order
 to facilitate the final design verification and its acceptance by certification authorities~\cite{dierkes2012}.
Our work addresses some of the issues encountered when attempting formal verification of properties of synchronous data flow models written in Lustre.
We propose a property-directed lemma generation approach, together with a prototype implementation. The proposed approach aims at reducing the amount of human
intervention usually needed to achieve {\em $k$-induction} proofs, possibly using {\em abstract interpretation} technique in cooperation.
Briefly outlined, the approach consists first in an abstract interpretation pass to discover coarse bounds on the numerical state variables of the
system; a $k$-induction engine and our lemma generation techniques are then ran in parallel
to search for potential invariants in order to strengthen the property.
We insist on the fact that the primary goal of the proposed method is discovering
missing information needed to prove properties the verification of which is either very expensive or impossible with currently available
methods and tools, rather than improving the performance of the verification of properties which are already relatively easily provable.

The paper is structured as follows: Section~\ref{systems} describes
the embedded software architectures targeted by our work. Related work and tools
are discussed in Section~\ref{related} before notations and vocabulary are
given in Section~\ref{notations}.
A description of the underlying $k$-induction engine assumed in this paper follows in Section~\ref{sec:kind}.
We introduce and motivate our approach in Section~\ref{approach} and detail the
lemma generation techniques in Section~\ref{hullqe}. The proposed approach is then
illustrated on a reconfiguration logic example and on Rockwell Collins industrial triplex sensor
voter in Section~\ref{applications}. Implementation is
briefly discussed in Section~\ref{framework}, before concluding in Section~\ref{conclusion}.

%% file: Systems.tex
\label{systems}

We consider embedded reactive software functions which contribute to
the safe operation of assemblies of hardware sensors, networked
computers, actuators, moving surfaces, {\em etc.} called
\emph{functional chains}. A functional chain can for instance be in
charge of "controlling the aircraft pitch angle", and must meet both
qualitative and quantitative safety requirements depending on the
effects of its failure. Effects are ranked from MIN (minor effect) to
CAT (catastrophic effect, with casualties). For instance, the failure of a pitch
control function is ranked CAT, and the function shall be robust to at least a double
failure and have an average failure rate of at most $\mathit{10^{-9}}$
per flight hour. In order to meet these requirements, engineers
must introduce hardware and software redundancy and implement several fault detection and
reconfiguration mechanisms in software. 

\begin{figure}[t]
\begin{center}
\scriptsize
\begin{tikzpicture}[scale=0.95]
\node[draw] (PITCH) at (-2,0) {PITCH};
\node[draw] (SPEED) at (-2,1) {SPEED};
\node[draw] (ORDER) at (-2,2) {ORDER};

\node[draw] (SMON3) at (-0.5,0) {SMON};
\node[draw] (SMON2) at (-0.5,1) {SMON};
\node[draw] (SMON1) at (-0.5,2) {SMON};

\node[draw] (VOTE13) at (1,0) {VOTE};
\node[draw] (VOTE12) at (1,1) {VOTE};
\node[draw] (VOTE11) at (1,2) {VOTE};

\node[draw] (LAW3) at (2.3,0) {LAW};
\node[draw] (LAW2) at (2.3,1) {LAW};
\node[draw] (LAW1) at (2.3,2) {LAW};

\node[draw] (VOTE23) at (3.6,0) {VOTE};
\node[draw] (VOTE22) at (3.6,1) {VOTE};
\node[draw] (VOTE21) at (3.6,2) {VOTE};

\node[draw] (AMON3) at (5,0) {AMON};
\node[draw] (AMON2) at (5,1) {AMON};
\node[draw] (AMON1) at (5,2) {AMON};

\node[draw] (RCF3) at (6.5,0) {RCF};
\node[draw] (RCF2) at (6.5,1) {RCF};
\node[draw] (RCF1) at (6.5,2) {RCF};

\node[draw] (ACT) at (8,1) {ACT};

\draw[->, >=latex] (PITCH) -- (SMON1);
\draw[->, >=latex] (SPEED) -- (SMON1);
\draw[->, >=latex] (ORDER) -- (SMON1);
\draw[->, >=latex] (SMON1) -- (VOTE11);
\draw[->, >=latex] (SMON1) -- (VOTE12);
\draw[->, >=latex] (SMON1) -- (VOTE13);
\draw[->, >=latex] (VOTE11) -- (LAW1);
\draw[->, >=latex] (LAW1) -- (VOTE21);
\draw[->, >=latex] (LAW1) -- (VOTE22);
\draw[->, >=latex] (LAW1) -- (VOTE23);
\draw[->, >=latex] (VOTE21) -- (AMON1);
\draw[->, >=latex] (ACT) -- (AMON1);
\draw[->, >=latex] (AMON1) -- (RCF1);
\draw[->, >=latex] (RCF1) -- (ACT);
\draw[->, >=latex] (RCF1) -- (RCF2);
\draw[->, >=latex] (RCF1) to[out=-50, in=50] (RCF3);

\draw[->, >=latex] (PITCH) -- (SMON2);
\draw[->, >=latex] (SPEED) -- (SMON2);
\draw[->, >=latex] (ORDER) -- (SMON2);
\draw[->, >=latex] (SMON2) -- (VOTE11);
\draw[->, >=latex] (SMON2) -- (VOTE12);
\draw[->, >=latex] (SMON2) -- (VOTE13);
\draw[->, >=latex] (VOTE12) -- (LAW2);
\draw[->, >=latex] (LAW2) -- (VOTE21);
\draw[->, >=latex] (LAW2) -- (VOTE22);
\draw[->, >=latex] (LAW2) -- (VOTE23);
\draw[->, >=latex] (VOTE22) -- (AMON2);
\draw[->, >=latex] (AMON2) -- (RCF2);
\draw[->, >=latex] (RCF2) -- (ACT);
\draw[->, >=latex] (ACT) to[out=-160,in=-20] (AMON2);
\draw[->, >=latex] (RCF2) -- (RCF3);

\draw[->, >=latex] (PITCH) -- (SMON3);
\draw[->, >=latex] (SPEED) -- (SMON3);
\draw[->, >=latex] (ORDER) -- (SMON3);
\draw[->, >=latex] (SMON3) -- (VOTE11);
\draw[->, >=latex] (SMON3) -- (VOTE12);
\draw[->, >=latex] (SMON3) -- (VOTE13);
\draw[->, >=latex] (VOTE13) -- (LAW3);
\draw[->, >=latex] (LAW3) -- (VOTE21);
\draw[->, >=latex] (LAW3) -- (VOTE22);
\draw[->, >=latex] (LAW3) -- (VOTE23);
\draw[->, >=latex] (VOTE23) -- (AMON3);
\draw[->, >=latex] (ACT) -- (AMON3);
\draw[->, >=latex] (AMON3) -- (RCF3);
\draw[->, >=latex] (RCF3) -- (ACT);
\end{tikzpicture}
\normalsize
\caption{Shuffled, triple channel architecture}
\label{fig:archi}
\end{center}
\vspace{-15pt}
\end{figure}
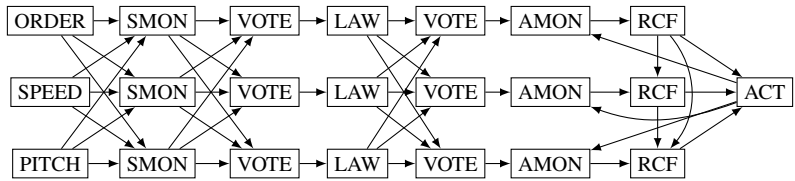

A frequently encountered architectural design pattern, triplication
with shuffle, is depicted in Fig.~\ref{fig:archi}.  It allows to recover from single failures and to detect double failures. Data sources, data
processing hardware and functions are triplicated to obtain three
\emph{channels}. The actuator is not replicated. Data is locally monitored
right after acquisition/production, but is also broadcast across
channels to be checked using triplex voting functions, in order to detect
complex error situations. Last, in each channel, depending on the
fault state of the channel and the observable behavior of other
channels, the \emph{reconfiguration logic} decides whether the channel in question
must take control of the actuator or on the contrary mute itself. Being \emph{healthy}
for a channel means that no fault has occurred for a sufficient amount
of consecutive time steps to become \emph{confirmed}.

Previous work by our team addresses the formal verification of
control laws numerical stability~\cite{RouxGaroche2011}, yet
ensuring proper behavior of \emph{voting functions} and
\emph{reconfiguration logic} as introduced here is equally important,
for these building blocks and design patterns are ubiquitous in fault
tolerant avionics software.  For the voting logic, we focus on BIBO
properties, ``bounded input implies bounded output'', the verification
of which is detailed in Section~\ref{applications:voter}. For
reconfiguration logic, which makes an extensive use of integer timers
and discrete logic, we focus on bounded liveness properties such as
``assuming at most two sensor, network or CPU faults, the actuator
must never remain idle for more than N consecutive time units'', the
verification of which is addressed in Section~\ref{applications:reconf}.

%% file: Context.tex
\label{related}

In this section we review the state of the art of verification tools
relevant in our application domain, \emph{i.e.}, of currently available
tools and techniques allowing to address synchronous data flow models
written in Lustre. We distinguish two main families of verification
approaches.

First, approaches based on abstract interpretation
(AI~\cite{DBLP:conf/popl/CousotC77}). The tool
NBac~\cite{DBLP:journals/fmsd/Jeannet03} for instance allows to
analyze properties of Lustre models by using a combination of forward
and backward fixpoint computation using AI. AI tends to need expert
tuning for the choice of abstract domains, partitioning, {\em etc.} to
behave correctly on the systems we consider. NBac proposes a heuristic
selection of AI parameter tuning, which dynamically refines domains
and partitionings to try to obtain a better precision without falling
in a combinatorial blowup.

Second, the family of $k$-induction~\cite{DBLP:conf/fmcad/SheeranSS00}
based approaches, with the commercial tool Scade Design
Verifier\footnote{\href{http://www.esterel-technologies.com/products/scade-suite/add-on-modules/design-verifier}{http://www.esterel-technologies.com/products/scade-suite/add-on-modules/design-verifier}},
or the academic tool Kind~\cite{DBLP:journals/corr/abs-1111-0372}. Kind is the most
recently introduced tool, and wraps the $k$-induction core in an
automatic counter example guided abstraction refinement loop whereas
the Scade Design Verifier does not.  $k$-induction is an exact
technique, in which little or no abstraction is performed (the
concrete semantics of the program is analyzed). Experiments show that
it does not scale up out of the box on the systems
encountered in our application field.  Proving proof obligations
on such systems often requires to unroll the system's transition relation to the
reoccurrence diameter of the model which can be very large in practice
(hundred or thousands of transitions).  For such proof
obligations, which are either $k$-inductive for a $k$ too large to be
reached in practice, or even non-inductive at all, numerical lemmas
are needed to help better characterize the reachable state space and
facilitate the inductive step of the
reasoning.

In order to address this common issue with $k$-induction, automatic
lemma generation techniques have been studied. Two main approaches can
be distinguished.  First, property agnostic approaches, such
as~\cite{KahGT-NFM-11}, in which template formulas are instantiated in
a brute force manner on combinations of the system state variables to
obtain a set of potential invariants. They are then analyzed alongside the
PO using the main $k$-induction engine.  Second, property directed approaches, such
as~\cite{DBLP:conf/ictac/BradleyM06,DBLP:conf/vmcai/Bradley11}, in
which the negation of root states of counterexamples are used as
strengthening lemmas, with or without generalization, or are used to
guide template instantiation. 
Also worth mentioning, interpolation~\cite{DBLP:conf/tacas/McMillan08}
yields very interesting results in lemma generation but unfortunately to our
knowledge no interpolation tool analyzing Lustre code exists.

We consider a lemma generation pass successful when the generated
potential invariants allow to prove the original proof objective with a
$k$-induction run with a small $k$. Once the right lemmas are found,
the proof can be easily re-run and checked by third party
$k$-induction tools, an important criterion for industrials and
certification organisms.
As we will see in the rest of the paper, the lemma generation approach proposed
in this paper takes inspiration from all the aforementioned techniques 
: while somehow brute force in its exploration of
the gray state space partitionings, our approach discovers relevant lemmas thanks
to its property-directed nature.


%% file: Notations.tex
\label{notations}

Let us now define several notions used throughout this paper. First, a
{\em transition system} is represented as a tuple $\langle v, D, I(v),
T(v,v')\rangle$ where $v$ is a vector of state variables, $D$
specifies the domain of each state variable, either boolean, integer
or real valued, $I$ is the initial state predicate, and $T$ is the
transition predicate in which $v'$ represents next state
variables. The logic used to express predicates is Linear
Integer or Real Arithmetic with Booleans. The usual
notions of trace semantics and reachability are used. Given a formula
$PO(v)$ representing a Proof Objective (PO), we say that the PO
holds if no state $s$ such that $\neg PO(s)$ can be reached
from $I$ through repeated application of $T$. Lustre or Scade programs
can be cast into this representation using adequate compilers.

An {\em atom} is a Boolean or its negation, or a linear equality or inequality in LRA or LIA.
A {\em polyhedron} is a conjunction of atoms.
More precisely, we will say polyhedron for not necessarily closed polyhedron,
meaning that we do not impose restrictions on the form of the inequalities
besides linearity.
The {\em convex hull} of two polyhedra $p_1$ and $p_2$ is the smallest polyhedron
such that it contains $p_1$ and $p_2$.
We will say that the convex hull $h$ of two polyhedra $p_1$ and $p_2$ is {\em exact}
if and only if $h=p_1\cup p_2$, and call it the {\em Exact Convex Hull} (or ECH) of
$p_1$ and $p_2$ if it exists.
For the sake of clarity, convex hulls that are not necessarily exact will be called
{\em Inexact Convex Hulls} (or ICH).
Note that for integer variables, the uniqueness of the convex hull is not guaranteed
if non-integer values for the coefficients are not forbidden.
We ban them in the rest of this paper; in our implementation, it is prevented by the type system.
Still, there are several ways to represent the same inequality, {\em e.g.}
$n>0$ and $n\geq1$.
Despite their difference in representation, these polyhedra
enclose the same (integer) points geometrically speaking, so this does not hinder our approach.
Convex hull comparison in this paper does not rely on their syntax nor semantics, but
rather on the {\em source} of the hull, {\em i.e.} the original polyhedra used
to create them.
This will be discussed in Section~\ref{hullqe} during the explanation
of our main contribution, the {\em hullification} algorithm.

%% file: kind.tex
\label{sec:kind}
The Stuff framework provides an SMT-based $k$-induction
module. Performing a $k$-induction analysis of a potential state invariant $P$
on a transition system $\langle I, T \rangle$ consists in checking the
satisfiability of the $\mathit{Base_k}(I, T, P)$ and
$\mathit{Step_k}(T,P)$ formulas, defined in \eqref{eq:kind}, for
increasing values of $k$, starting from a user specified $k>1$.

\begin{equation}
  \label{eq:kind}
  \begin{split}
    \mathit{Base_k}(I, T, P) & \equiv  
    \overbrace{I(s_0)}^{\text{Initial state}} \land
    \overbrace{\bigwedge_{i \in [0,k-2]}{T(s_i,s_{i+1})}}^{\text{trace of k-1 transitions}} \land
    \overbrace{\bigvee_{i \in [0,k-1]} \lnot P(s_i)}^{\text{$P$ falsified on some state}} \\
    \mathit{Step_k}(T,P) & \equiv
    \underbrace{\bigwedge_{i \in [0,k-1]} T(s_i,s_{i+1})}_{\text{trace of $k$ transitions}} \land
    \underbrace{\bigwedge_{i \in [0,k-1]} P(s_i)}_{\text{$P$ satisfied on first $k$ states}} \land
    \underbrace{\lnot P(s_k)}_{\text{$P$ falsified by last state}}
  \end{split}
\end{equation}

The base and step instances are analysed, until either a base model
has been found, in which case the proof objective is falsified, a user
specified upper bound for $k$ has been reached for base and step, in
which case the status of the proof objective is still undefined, or a
$k$ value has been discovered so that both formulas are unsatisfiable,
which proves the validity of the objective.

In addition, this $k$-induction engine allows, for any $n$, to partition a given set
of proof objectives $P = \{P_j\}$, viewed as a conjunction
$P = \bigwedge_j{P_j}$, in three maximal subsets $F_n$, $U_n$ and $V_n$, such
that: 

\begin{itemize}
\item elements $P \in F_n$ are such that
  $\mathit{Base_n(I,T,P)}$ is satisfiable: they are \emph{Falsified};
\item elements $P \in U_n$ are such that $\mathit{Base_n(I,T,P)}$ is
  unsatisfiable and $\mathit{Step_n(T,P)}$ is satisfiable: they are
  \emph{Undefined} because neither falsifiable nor $n$-inductive;
\item elements of $V_n$ are such that
  $\mathit{Base_n(I,T,\bigwedge_{P \in V_n}{P})}$ is unsatisfiable and
  $\mathit{Step_n(T,\bigwedge_{P \in V_n}{P})}$ is unsatisfiable: they
  are mutually $n$-inductive, \emph{i.e.} \emph{Valid} on the transition
  system.
\end{itemize}

%% file: Approach.tex
\label{approach}
Our lemma generation heuristic builds on a backward property-directed
reachability analysis.
We use Quantifier Elimination (QE~\cite{DBLP:conf/lpar/Monniaux08,Nipkow10,bjorner10}) to compute
successive preimages of the negation of the PO, in the spirit
of~\cite{DBLP:conf/cav/MouraRS03,DBLP:conf/atva/DammDHJPPSWW07}. In
our approach, the states characterized by the preimages are generated
in a way such that
\begin{inparaenum}[\itshape (i)]
\item they satisfy the PO and
\item from them, it is possible to reach a state violating the PO if certain transitions are taken.
\end{inparaenum}
Such states will be referred to as \emph{gray states}.
This can be achieved by calculating the preimages as follows:
\begin{equation}
\label{eq:preimage}
\begin{split}
\mathit{preimage_1} &= \mathit{QE}(s',\mathit{PO}(s) \land T(s,s') \land \neg\mathit{PO}(s'))\\
\mathit{preimage_i} &= \mathit{QE}(s',\mathit{PO}(s) \land T(s,s') \land \mathit{preimage_{i-1}}[s'/s]) \quad( \text{for}\;i>1 )
\end{split}
\end{equation}
where $QE(\vec{v},F)$ returns a quantifier-free formula equisatisfiable to $\exists\vec{v},\;F$ and such that $FV(QE(v,F))=FV(F)\setminus v$.
The preimages themselves are assumed to be in DNF, by using
\cite{DBLP:conf/lpar/Monniaux08} as a QE engine for instance.

From these preimages we extract information using two search
heuristics introduced and motivated in the rest of this section and
detailled in Section~\ref{hullqe}. These heuristics run in parallel,
alongside the backward analysis computing the next preimage and a
$k$-induction engine.  The backward analysis is not run to a fixed
point before proceeding further, it is rather meant to probe the gray
state space around the negation of the PO, and feeding the potential lemma generation with the preimages
as soon as they are produced.

To extract information out of the preimages, at any point of the backward exploration, their disjunction is considered: it represents
the gray states found so far as a union of polyhedra. The main idea underlying the potential lemma generation
is to explore the ways in which
those polyhedra can be grouped using convex hull calculation, thus discovering linear relations over state variables
representing boundaries between convex regions of the gray state space.  
Since these convex boundaries enclose unauthorized states, they are negated before
being sent to our $k$-induction engine to check their validity and try to stenghten the PO.

The PO is successfully strengthened by a set of lemmas when the
set $V_k$  of valid POs, produced by the $k$-induction analysis detailled in Section~\ref{sec:kind},
contains the main PO at the end of a run.
If the original PO is not strengthened by the potential invariants extracted from the
currently available preimages, a new preimage is calculated, bringing more information.
Yet, when the PO is strengthened and proved valid, it can be the case that not all elements
of the valid subset $V_k$ are needed to
entail the original PO. A minimization pass inspects them one by one,
discarding $l\not=PO$ from $V_k$ if
$\bigwedge(V\setminus l)$ remains
$k$-inductive, to obtain a relatively small and readable set of lemmas.

Note that, in the backward exploration, the choice of which variables to eliminate by QE and which to keep is important.
Eliminating the next state variables
and keeping the current state variables is not satisfactory in the general case, as on large scale systems, many state variables might not be relevant for the PO under
investigation, and might hinder the performance of the convex hull calculation or $k$-induction.
Therefore, the only state variables that are \textbf{not} eliminated are the ones found
in the cone of influence of the PO, in their {\em current state} version. In particular, the system inputs are eliminated since they do not provide more information
from a backward analysis point of view.

\begin{wrapfigure}{r}{.55\textwidth}
  \vspace{-10pt}
  \begin{subfigure}[b]{.2\textwidth}
    \begin{center}
      \scriptsize
      \begin{tikzpicture}[scale=.5]
        \draw [->] (0,0) -- coordinate (x axis mid) (4.5,0) node[right] {$x$};
        \draw [->] (0,0) -- coordinate (y axis mid) (0,3.5) node[left] {$y$};
        \foreach \x in {0,...,4}
        \draw (\x,1pt) -- (\x,-3pt)
        node[anchor=north] {\x};
        \foreach \y in {0,...,3}
        \draw (1pt,\y) -- (-3pt,\y) 
        node[anchor=east] {\y};
        \draw (0,0) -- (1,1);
        \draw (0,0) -- (1,0);
        \draw (1,0) -- (1,1);
        \draw (2,0) -- (2,1);
        \draw (2,1) -- (3,1);
        \draw (3,1) -- (3,0);
        \draw (2,0) -- (3,0);
        \draw (3,2) -- (3,3);
        \filldraw[black] (0,0) circle (2pt);
        \filldraw[black] (1,0) circle (2pt);
        \filldraw[black] (2,0) circle (2pt);
        \filldraw[black] (3,0) circle (2pt);
        \filldraw[black] (4,0) circle (2pt);
        \filldraw[black] (1,1) circle (2pt);
        \filldraw[black] (2,1) circle (2pt);
        \filldraw[black] (3,1) circle (2pt);
        \filldraw[black] (2,2) circle (2pt);
        \filldraw[black] (3,2) circle (2pt);
        \filldraw[black] (3,3) circle (2pt);
        \draw[dashed] (1,1) -- (3,3);
        \draw[dashed] (3,3) -- (4,0);
        \draw[dashed] (1,0) -- (2,0);
        \draw[dashed] (3,0) -- (4,0);
        \draw[dotted] (3,1) -- (4,0);
        \draw (0.75,0.37) node {$s_1$};
        \draw (2.5,0.5)   node {$s_2$};
        \draw (1.5,2.25)  node {$s_3$};
        \draw (3.5,3.25)  node {$s_4$};
        \draw (4.5,0.25)  node {$s_5$};
      \end{tikzpicture}
    \end{center}
    \caption{ECH on integers}\label{fig:echInt}
  \end{subfigure}
  ~
  \begin{subfigure}[b]{.3\textwidth}
    \begin{center}
      \scriptsize
      \begin{tikzpicture}[scale=.8]
        \draw [->] (0,0) -- coordinate (x axis mid) (4.5,0) node[right] {$x$};
        \draw [->] (0,0) -- coordinate (y axis mid) (0,2.5) node[left] {$y$};
        \foreach \x in {0,...,4}
        \draw (\x,1pt) -- (\x,-3pt)
        node[anchor=north] {\x};
        \foreach \y in {0,...,2}
        \draw (1pt,\y) -- (-3pt,\y) 
        node[anchor=east] {\y};
        \draw (0,0) -- (0,2);
        \draw (0,0) -- (3.85,0);
        \draw (0,2) -- (1.85,2);
        \draw (3.85,0) -- (1.85,2);
        \draw (2,2) -- (4,0);
        \draw (1,1)   node {$s_1$};
        \draw (3.5,1) node {$s_2$};
      \end{tikzpicture}
    \end{center}
    \caption{ECH on Reals}\label{fig:echReal}
  \end{subfigure}
  \caption{New relations with hulls}
  \label{fig:polys}
  \vspace{-10pt}
\end{wrapfigure}
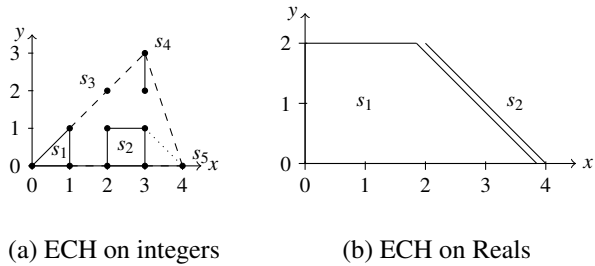

Before going into the details of the potential lemma generation
algorithm, let us illustrate how computing ICHs and ECHs can actually
make new numerical relations appear, using the
examples given in Figure~\ref{fig:echInt} and
Figure~\ref{fig:echReal}.

In Figure~\ref{fig:echInt}, the gray state space of a system with two integer state variables is represented.
States are represented as dots, polyhedron $s_1$ contains three
states, polyhedra $s_3$ and $s_5$ only contain one state {\em etc}.

Computing exact convex hulls over these base polyhedra in the LIA fragment yields (at least) two new borders,
{\em i.e.} potential relational invariants, pictured as dashed lines.
An example of merging order is to merge $s_1$ with $s_2$, $s_3$ with $s_4$, $\{s_1,s_2\}$
with $\{s_3,s_4\}$, and $\{s_1,s_2,s_3,s_4\}$ with $s_5$ $(1)$.

On a system with real valued state variables however, as shown in Figure~\ref{fig:echReal}, the only case in which we will {\em discover} a new border
by computing exact convex hulls is when one is the limit of another, as illustrated on Figure~\ref{fig:echReal}.
Here $s_1$ is made of $0\leq x$, $0\leq y\leq2$ and $y+x-4<0$;
$s_2$ is made of $0\leq y\leq2$ and $y+x-4=0$, so the resulting hull will be
$0\leq x$, $0\leq y\leq2$ and $y+x-4\leq0$. The information learned this way has little chance
of strengthening the PO.

As will be seen in the next sections, when trying to discover new relations, ECH-based techniques work best
for integer valued systems, while ICH can be beneficial for both real or integer valued systems.

\subsection{A First Example}
\label{approach:counter}

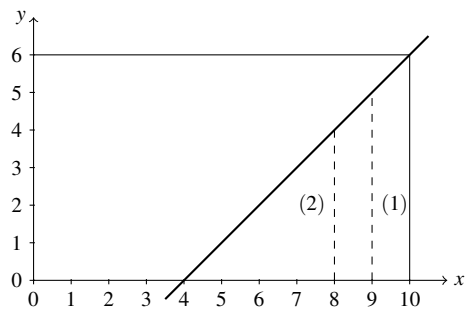
\begin{wrapfigure}{r}{.5\textwidth}
  \vspace{-20pt}
  \scriptsize
  \begin{center}
    \begin{tikzpicture}[scale=.5]
      \draw [->] (0,0) -- coordinate (x axis mid) (11,0) node[right] {$x$};
      \draw [->] (0,0) -- coordinate (y axis mid) (0,7) node[left] {$y$};
      \foreach \x in {0,...,10}
      \draw (\x,1pt) -- (\x,-3pt)
      node[anchor=north] {\x};
      \foreach \y in {0,...,6}
      \draw (1pt,\y) -- (-3pt,\y) 
      node[anchor=east] {\y};
      \draw (0,6) -- (10,6);
      \draw (10,0) -- (10,6);
      \draw [thick] (3.5,-0.5) -- (10.5,6.5);
      \draw [dashed] (9,0) -- (9,5);
      \draw (9.6,2) node {$(1)$};
      \draw [dashed] (8,0) -- (8,4);
      \draw (7.4,2) node {$(2)$};
    \end{tikzpicture}
  \end{center}
  \vspace{-10pt}
  \caption{ECH calculation on the double counter}
  \label{fig:dblc}
\end{wrapfigure}

We consider a simple example called the double counter\footnote{Code available
at~\href{http://www.onera.fr/staff-en/adrien-champion/}{http://www.onera.fr/staff-en/adrien-champion/}.} with two integer state
variables $x$ and $y$ and three boolean inputs $a$, $b$ and $c$. Variables $x$ and $y$ are
initialized to $0$, and are both incremented by one when $a$ is true or keep their
current value when $a$ is false. The variable $x$ is reset if $b\lor c$ is true,
and saturates at $n_x$. The variable $y$ is reset when $c$ is true and saturates at $n_y$,
hence $y$ cannot be reset without resetting $x$, and $n_x>n_y$.
The proof objective is $x=n_x\Rightarrow y=n_y$. 
Here is a possible transition relation for such a system:
\begin{center}
\small
\begin{tabular}{r c l l l l}
  $ T(s,s') =$ &         & $\big(\text{if}\;(b\lor c)$ & $\text{then}\;x'=0\;\text{else if}\;(a\land x<n_x)$ & $\text{then}\;x'=x+1$ & $\text{else}\;x'=x\big)$\\
               & $\land$ & $\big(\text{if}\;(c)$ & $\text{then}\;y'=0\;\text{else if}\;(a\land y<n_y)$ & $\text{then}\;y'=y+1$ & $\text{else}\;y'=y\big)$.\\
\end{tabular}
\end{center}

Let us see now how the proposed approach performs on this system
when fixing $n_x = 10$ and $n_y = 6$ for instance.
First, using the abstract interpretation tool presented in~\cite{DBLP:journals/entcs/RouxDG10},
bounds on $x$ and $y$ are easily discovered: $0\leq x\leq n_x = 10$ and
$0\leq y\leq n_y = 6$, yet the PO cannot be proved with AI without further manual intervention.
So, using these range properties once $k$-induction has confirmed them, we start the
backward property-directed analysis, which outputs a first preimage:
$x = 9\;\land\;0 \leq y < 5\;(1)$.
Unsurprisingly, it is too weak to conclude, {\em i.e.} its negation is not $k$-inductive for a small $k$.
The next preimage is $x = 8\;\land\;0 \leq y < 4\;\lor x=9\;\land\;0\leq y<5\;(2)$ which
does not allow to conclude either for the same reason.
Instead of iterating until a fixed point is found, consider the graph on Figure~\ref{fig:dblc}.
It shows the two first preimages as dashed lines which seem to suggest a relation between $x$ and
$y$, pictured as a bold line.
This relation can be made explicit by calculating the convex hull of the disjunction of the first
two preimages -- since this particular system can stutter, it is the same as $(2)$.
This yields $8 \leq x \leq 9 \;\land\; 0 \leq y < x - 4$.
Note that this convex hull is an ECH, since both $x$ and $y$ are integers.
The four inequalities are negated -- they characterize gray states --
and are sent to the $k$-induction engine.
Potential invariants $\neg 8\leq x$, $\neg x\leq 9$ and $\neg 0 \leq y$ are falsified,
and the PO in conjunction with lemma $\neg y < x-4$ is found to be 1-inductive.

In fact, this PO could also be proved correct by $k$-induction given the bounds found by AI only,
by unrolling the transition relation to the reoccurrence diameter of the system.
In practice, even on such a simple system it is not possible for large
values of $n_x$ and $n_y$ (hundreds or thousands of transitions).
The performance of our technique on the other hand is not sensitive to the actual value of numerical constants: it will always
derive the strengthening lemma from the first two preimages.
Obviously, the time needed to compute the preimages is not impacted by changing the constants values either.\\
For more complex systems with preimages made of more than two polyhedra, simply merging
them in arbitrary order using convex hull calculation is not robust since
the resulting convex hulls would depend on the merging order, and
interesting polyhedra could be missed.
This idea of an exhaustive enumeration of the intermediary ECHs that can appear
when merging a set of polyhedra is explored in Section~\ref{hullqe:algorithm}.

\subsection{A Second Example}
\label{approach:duplex}

\begin{figure}[t]
  \begin{center}
  \begin{subfigure}[b]{.4\textwidth}
    \begin{center}
      \includegraphics[width=\textwidth]{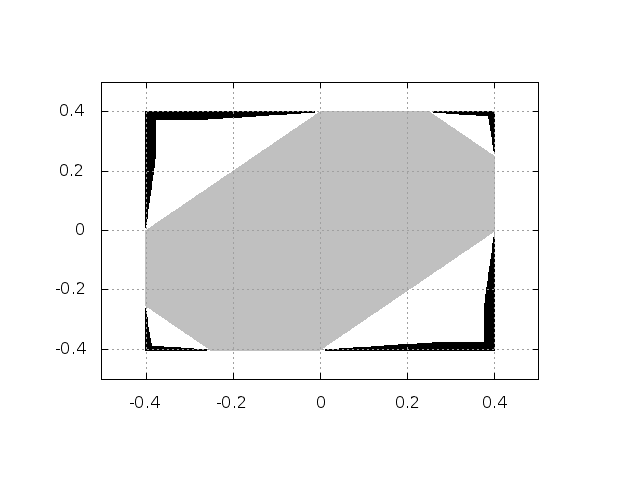}
    \end{center}
    \vspace{-20pt}
    \caption{Two inputs voter, first preimage}\label{fig:duplex1}
  \end{subfigure}
  \begin{subfigure}[b]{.4\textwidth}
    \begin{center}
      \includegraphics[width=\textwidth]{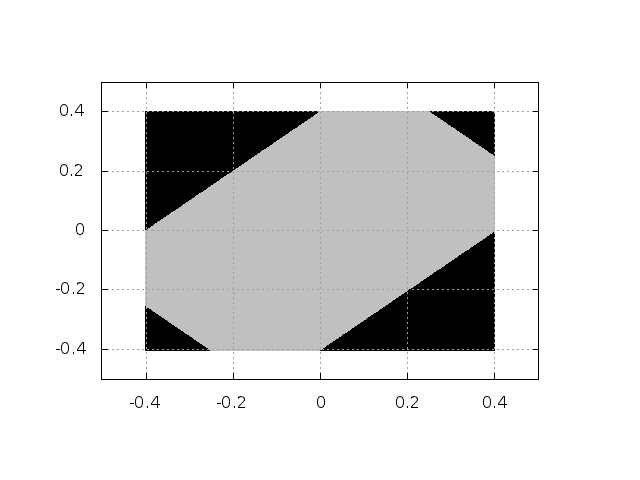}
    \end{center}
    \vspace{-20pt}
    \caption{Two inputs voter with ICH}\label{fig:duplex2}
  \end{subfigure}
  \end{center}
  \vspace{-10pt}
  \caption{Simple voting logic.}
  \label{fig:duplex}
  \vspace{-10pt}
\end{figure}

Let us now consider briefly a two input, real valued voting logic system
derived from the Rockwell Collins triplex voter.
We will not discuss the system itself since the triplex
voter is detailled in Section~\ref{applications:voter}. It simply allows us to represent
graphically the state space in a plan.
The PO here is that two of the state variables, $Equalization_1$ and $Equalization_2$, range
between $-0.4$ and $0.4$.
Figure~\ref{fig:duplex} depicts the corresponding square.
On Figure~\ref{fig:duplex1} we can see the first preimage calculated by our backward reachability
analysis as black triangles, and the strengthening lemmas found by hand in~\cite{dierkes11} transposed
to the two input system as a gray octagon.
Calculating ECH on this first preimage does not allow to conclude.

A more relevant approach would be to calculate ICH. Yet, since the ICH of all the preimage polyhedra
is the $[-0.4,0.4]^2$ square, we need to be more
subtle and introduce a criterion for ICH to be actually computed between two polyhedra: they have
to intersect. Intersection can be checked by a simple satisfiability test performed using a SMT solver.
This check allows us to identify overlapping areas of the gray state space and to over-approximate them, while not merging disjoint areas in the gray state space explored so far.
This approximation obtained through ICH resembles widening techniques used in abstract
interpretation~\cite{DBLP:conf/popl/CousotC77} in the sense that it allows to {\em jump} forward in the
analysis iterations, yet it differs in the sense that, contrary to widening, it does not ensure
termination. The only goal here is to generate potential invariants for the PO, and Figure~\ref{fig:duplex} shows that the
ICH yields exactly the dual, in the $[-0.4,0.4]^2$ square, of the octagon invariant found by hand
in~\cite{dierkes11}. 
This second idea of using ICHs to perform overapproximations will be discussed in Section~\ref{hullqe:ich}.

%% file: Hullqe.tex
\label{hullqe}

We now detail two heuristics which use the preimages output by the backward analysis.
The first one follows the example from Section~\ref{approach:counter} and consists
in a thorough, exact exploration of the partitionings of the gray state space.
After explaining the basic algorithm in Section~\ref{hullqe:algorithm}, optimizations are
developed in Section~\ref{hullqe:optimization}. A small example illustrates
the method in Section~\ref{hullqe:example}.
The second heuristic over-approximates areas of the gray state space in the spirit of the
discussion in Section~\ref{approach:duplex}, and is discussed
in Section~\ref{hullqe:ich}.
Both aim at discovering new relations between the state variables which once negated
become potential invariants.
Figure~\ref{fig:global} provides a high level view of the different components and the way
they interact internally and with the exterior.

\subsection{Hullification Algorithm}
\label{hullqe:algorithm}

The algorithm presented in this section, called \emph{hullification}, calculates all
the convex hulls that can be
created by iterating the convex hull calculation on a given set of polyhedra, called
the {\em source} polyhedra.
In this algorithm we will calculate ECH as opposed to ICH to avoid both losing precision
in the process and the potential combinatorial blow up -- ICH are used in a different
approach in Section~\ref{hullqe:ich}.
The difficulty here is to not miss any of the ECH that can be possibly
calculated from the source polyhedra. Indeed, back to the example on Figure~\ref{fig:echInt}
the merging order $(1)$
misses the ECH of $s_2$ and $s_5$ (represented as a dotted line), and consequently the potential relational 
lemma $y\leq-x+4$, which could have strengthened the PO.

Imperative and slightly object-oriented pseudo-code is provided on
Algorithm~\ref{alg:hull}.
The purpose of $generatorSetMemory$ is related to optimizations,
discussed in Section~\ref{hullqe:optimization}.
Please note that for the sake of clarity, the function called on line~$20$ is
detailed separately on Algorithm~\ref{alg:update}.
The hullification algorithm iterates on a set of pairs called the $generatorSet$:
the first component of each of these pairs is a convex hull called the {\em pivot}.
The second one is a set of convex hulls the pivot will be tried to be exactly merged with,
called the pivot {\em seeds}.
Note that since the ECHs are calculated by merging polyhedra two by two,
our hullification algorithm cannot find convex hulls that require to merge more than
two polyhedra at the same time to be exact.\\
The $generatorSet$ is initialized such that for any couple $(i,j)$ such that $0\leq i\leq n$
and $i< j\leq n$, $p_i$ is a pivot and $p_j$ is one of its seeds, line~$3$.
A $newgeneratorSet$ is initialized with the same
pivots as the $generatorSet$ but without any seeds (line~$8$).
At each iteration (line~$6$), a first loop enumerates the pairs of pivot and seeds
of the generation set (line~$9$).
Embedded in the first one, a second loop iterates on the seeds (line~$11$) and tries
to calculate the ECH of the pivot and the seed (line~$14$) as described in
Section~\ref{notations}.
If the exact merge was successful, the new ECH is added to the seeds of the pivots of
the $newGeneratorSet$ (line~$20$, detailled below) and as a new pivot with no seeds.
Once the elements of the $generatorSet$ have all been inspected and if new
ECH(s) have been found, a new iteration begins with the $newGeneratorSet$.
When no new convex hulls are discovered during an iteration, the algorithm returns all the ECHs found so far (line~$27$).

\begin{algorithm}[t!]
\scriptsize
\caption{Hullification Algorithm:\\$hullification(\{p_i|0\leq i\leq n\})$.}
\label{alg:hull}
\begin{algorithmic}[1]
\State $generatorSetMemory=\{\{p_i\}|0\leq i\leq n\}$
\State $sourceMap=\{p_i\rightarrow \{p_i\}|0\leq i\leq n\}$
\State $generatorSet=\{(p_i,S_i)|0\leq i\leq n\land S_i=\{p_k|i<k\leq n\}\}$
\State $generatorSetMemory = generatorSetMemory \cup \{\{p_i,p_j\}|0\leq i\leq n,\; i< j\leq n\}$
\State $fixedPoint=\mathbf{false}$
\While {$(\neg\text{fixedPoint})$}
  \State $fixedPoint=\mathbf{true}$
  \State $newGeneratorSet=\{(p_i,\{\})|\exists S,(p_i,S)\in generatorSet\}$
  \ForAll {$((pivot,seeds)\in generatorSet)$}
    \State $sourcePivot=sourceMap.get(pivot)$
    \ForAll {$(seed\in seeds)$}
      \State $sourceSeed=sourceMap.get(seed)$
      \State $source=sourcePivot\cup sourceSeed$
      \State $hull=computeHull(pivot,seed)$
      \State $newGeneratorSet.update(pivot,newGeneratorSet.get(pivot) - seed)$
      \If {$(hull\not=\mathbf{false})$}
        \State $fixedPoint=\mathbf{false}$
        \State $sourceMap.add(hull\rightarrow source)$
        \State $newGeneratorSet=$
        \State $\quad\quad updateGenSet(hull,source,pivot,seed,newGeneratorSet)$
      \EndIf
    \EndFor
  \EndFor
  \State $generatorSet=newGeneratorSet$
  \State // Communication.
\EndWhile
\State $\mathbf{return}\; \{p_i|\exists S,(p_i,S)\in generatorSet\}$
\end{algorithmic}
\end{algorithm}

\begin{algorithm}[t!]
\scriptsize
\caption{Updating the $newGeneratorSet$:\\$updateGenSet(hull,source,newGeneratorSet)$.}
\label{alg:update}
\begin{algorithmic}[1]
\State $result=\{\}$
\ForAll {$((pivotAux,seedsAux)\in newGeneratorSet)$}
  \State $sourceAux = sourceMap.get(pivotAux)$
  \State $shallAdd = (sourceAux\cup source)\not\in generatorSetMemory\;\&\&$
  \State $\quad\quad\quad\quad\quad(sourceAux \not\subset source)$
  \If {$(shallAdd)$}
    \State $result.update((pivotAux,seedsAux\cup\{hull\}))$
    \State $generatorSetMemory.add(sourceAux\cup source)$
  \Else
    \State $result.add((pivot,seeds))$
  \EndIf
\EndFor
\State $result.add((hull,\{\}))$
\State $\mathbf{return}\;result$
\end{algorithmic}
\end{algorithm}

\subsection{Optimizing Hullification}
\label{hullqe:optimization}

The hullification algorithm is highly combinatorial, and this section
presents optimizations that improve its scalability.


In the hullification algorithm, the number of merge attemps increases
dramatically depending on the number of elements added in the
$generatorSet$ at each iteration. With hullification as is, in many
cases, elements of this set can be redundant, in the sense that the
new hulls derived from them, if any, would be the same even though the
elements are different. The key idea to reducing redundancy is to
keep a link between any ECH calculated and the source polyhedra merged
to create it, thereafter called the ECH source, and use this
information to skip redundant ECH calculation attempts.

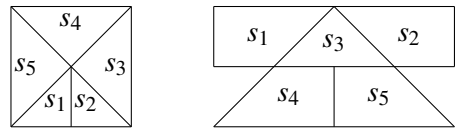
\begin{wrapfigure}{R}{.45\textwidth}
  \vspace{-10pt}
  \begin{center}
  \begin{subfigure}[b]{.2\textwidth}
    \begin{center}
      \begin{tikzpicture}[scale=.8]
        \draw (0,0) -- (2,0);
        \draw (0,0) -- (0,2);
        \draw (2,0) -- (2,2);
        \draw (0,2) -- (2,2);
        \draw (0,0) -- (2,2);
        \draw (0,2) -- (2,0);
        \draw (1,0) -- (1,1);
        \draw (0.75,0.4) node {$s_1$};
        \draw (1.25,0.4) node {$s_2$};
        \draw (1.75,1)   node {$s_3$};
        \draw (1,1.75)   node {$s_4$};
        \draw (0.25,1)   node {$s_5$};
      \end{tikzpicture}
    \end{center}
    \caption{Square example}\label{fig:sqr}
  \end{subfigure}
  ~
  \begin{subfigure}[b]{.2\textwidth}
    \begin{center}
      \begin{tikzpicture}[scale=.8]
        \draw (0,0) -- (2,2);
        \draw (0,0) -- (4,0);
        \draw (2,0) -- (2,1);
        \draw (2,2) -- (4,0);
        \draw (0,1) -- (4,1);
        \draw (0,2) -- (4,2);
        \draw (0,2) -- (0,1);
        \draw (4,2) -- (4,1);
        \draw (0.75,1.5) node {$s_1$};
        \draw (3.25,1.5) node {$s_2$};
        \draw (2,1.4)    node {$s_3$};
        \draw (1.25,0.5) node {$s_4$};
        \draw (2.75,0.5) node {$s_5$};
      \end{tikzpicture}
    \end{center}
    \caption{Hat example}\label{fig:hat}
  \end{subfigure}
  \vspace{-10pt}
  \end{center}
  \caption{Hullification redundancy issues}
  \label{fig:polys}
  \vspace{-8pt}
\end{wrapfigure}

Consider for example Figure~\ref{fig:sqr}.  If we already tried to
merge the ECH of source $\{s_1,s_2,s_3\}$ with the one of source
$\{s_4,s_5\}$ then it is not necessary to consider trying to merge say
the ECH of source $\{s_3,s_4\}$ with the one of source
$\{s_1,s_2,s_5\}$.  The result would be the same, {\em i.e.} the same
ECH or a failure to merge the convex hulls exactly (the same ECH here).
Note that since
we are generating all the existing ECHs from the source polyhedra,
this case happens every time an ECH can be calculated by merging its
source in strictly more than one order, that is to say \textbf{very}
often. More generally, we do not want to attempt merges of different
hulls deriving from the same set of source polyhedra.

Another source of redundancy is that, when a seed is added to a pivot during the $generatorSet$ update,
it represents a potential merge of the union of the pivot source and the seed source.
Even if this merge has not yet been considered, a potential merge of the same source might
have already been added to the $generatorSet$ through a different seed added to a different
pivot. In this case we do not want the seed to be added.
So, in order to prevent redundant elements from being added to the $generatorSet$,
we introduce a memory called $generatorSetMemory$, and
control how new hulls are added to the $newGeneratorSet$.
For a new hull to be added to a pivot as a seed,
$source(pivot)\cup source(hull)\not\in generatorSetMemory$ must hold
(Algorithm~\ref{alg:update} line~$4$); if the hull is indeed added to
the seeds of the pivot, then $generatorSetMemory += source(pivot)\cup source(hull)$
(Algorithm~\ref{alg:update} line~$8$).  Informally, this memory contains the
sources of all the potential merges added to the generator set. This
ensures that the merge of a source will never be considered more than
once, and that
the merges we did not consider were not reachable by successive pair-wise ECH calculation.

Also, we forbid adding a seed to a pivot's seeds if the source of
the latter is a subset of the former, since the result would
necessarily be
the seed itself (we call this $(1)$).
Another improvement deals with \emph{when} hullification interacts with the the rest
of the framework. Since our goal is to generate potential invariants,
we do not need to wait for the hullification algorithm to terminate to
communicate the potential invariants already found so far.  They are
therefore communicated, typically to $k$-induction, after each {\em
  big iteration} of the algorithm (loop on Algorithm~\ref{alg:hull}
line~$25$).  This has the added benefit of launching $k$-induction on
smaller potential invariant sets.  

There is a drawback in comparing hulls using their sources:
assume that two of the input (source) polyhedra $p_i$ and $p_j$ are such that
$p_i\Rightarrow p_j$.
Then the exact merge of $p_i$ and $p_j$ succeeds and yields the hull of source
$\{p_i,p_j\}$, which is really $p_j$.
As a consequence, the pivots of the $generatorSet$ are redundant, as are their seeds
and in the end the merge attempts.
To avoid this, we first check the set of input polyhedra and discard redundant ones.

Last but not least, merges are also memorized in between calls to the
algorithm so that we do not call the merge algorithm when considering two
polyhedra we already merged during a previous call. Since
hullification is called on the ever-growing disjunction of all
preimages found so far, each new disjunction contains the previous
one and this represents a significant improvement.

In the next subsection we illustrate hullification on a small example before introducing
another potential invariant generation algorithm in Section~\ref{hullqe:ich}.
Hullification will be illustrated on a reconfiguration logic system in Section~\ref{applications:reconf}

\subsection{Hullification Example}
\label{hullqe:example}

Let us now unroll the algorithm on a simple example depicted on Figure~\ref{fig:hat}.
For the sake of concision a source $\{s_1,s_2,\cdots,s_n\}$ will be written $12\cdots n$.
We write generator sets in the following fashion: $\{(pivot,[seeds])\}$.\\
With this convention, the initial $generatorSet$ is $\{(1,[2,3,4,5]),(2,[3,4,5]),(3,[4,5]),(4,[5]),(5,[])\}$.
The $newGeneratorSet$ for the first {\em big step} iteration trace is as follows:

\begin{center}
\begin{tabular}{l l l l l l l l}
  $1,[]$ & $2,[]$ & $3,[]$ & $4,[]$ & $5,[]$ & & & \\\hline
  $1,[]$ & $2,[13]$ & $3,[]$ & $4,[13]$ & $5,[13]$ & $13,[]$ & & \\\hline
  $1,[]$ & $2,[13]$ & $3,[]$ & $4,[13,23]$ & $5,[13,23]$ & $13,[]$ & $23,[]$ & \\\hline
  $1,[45]$ & $2,[13,45]$ & $3,[45]$ & $4,[13,23]$ & $5,[13,23]$ & $13,[45]$ & $23,[45]$ & $45,[]$ \\
\end{tabular}
\end{center}

At first $newGeneratorSet$ is the same as $generatorSet$ without seeds (first line of the trace).
We first consider $1$ as a pivot. The merge of $1$ and $3$ works while the other ones fail, leading to the
second line of the trace.
Note that $13$ is not added to $1$ nor $3$ since $1\subseteq 13$ and $3\subseteq 13$ by $(1)$.
With this pivot we add three sources to the $generatorSetMemory$: $213$, $413$ and $513$ $(2)$.
The next pivot is $2$ which is merged with $3$ while the merges with the other
seeds fail.
After the $newGeneratorSet$ update we obtain the third line of the trace.
Note that $23$ is not added to the seeds of $1$ since source $213$ has already been added to the
$generatorSetMemory$ at $(2)$ so $23\cup 1\in generatorSetMemory$.
Similarily, it is not added to the seeds of $13$ either.
Next pivot $3$ cannot be merged with any of its seeds.
Pivot $4$ can be merged with $5$ producing the fourth line of the generator trace.
A new big step iteration begins during which $2$ will be merged with $13$ and $3$ with $45$
while all the other merges will fail. At the beginning of the third big step iteration the
$generatorSet$ is \[\{(1,[345]),(2,[345]),(3,[]),(4,[123]),(5,[123]),\]\[(13,[345]),(23,[345]),(45,[123]),(123,[345]),(345,[])\}.\]
No new hull is found and the algorithm detects that a fixed point has been reached.

\subsection{Another Way to Generate Potential Invariants: ICHs}
\label{hullqe:ich}

As mentioned before in Section~\ref{hullqe:algorithm}, ECH calculation cannot do much
for real state variables.
We therefore propose a second approach based on Inexact Convex Hull (ICH) calculation
modulo intersection as mentioned in Section~\ref{approach:duplex},
simply called ICH calculation in the rest of this paper.
That is, two polyhedra will be inexactly merged if and only if their intersection
is not empty.
This regroups areas of the gray state space that are not disjoint and over-approximates them
to make new numerical relations appear.
An efficient way to check for intersection is to check the satisfiability of the
conjunction of the constraints describing the two polyhedra using an SMT solver.
Note that this technique is also of interest in the integer case.

For a given set of polyhedra more ICHs than ECHs can be created, in practice often
a lot more. The hullification algorithm using ICHs thus tends to choke.
We propose the following algorithm, only briefly described for the sake of concision.\\
Select a pivot in the input polyhedra set and try to find an ICH with the other ones.
If an ICH with another polyhedra (source) exists, both the pivot and the source are
discarded, and the ICH becomes the new pivot.
Once all the merges have been tried, the pivot is put aside and a new pivot is selected
in the remaining polyhedra set.
When the algorithm runs out of polyhedra, it starts again on the polyhedra put aside
if at least one new hull was found.
If not, a fixed point has been reached and the algorithm stops.\\
Although the intermediary ICHs computed in this algorithm depends on the order in which
the pivots are selected and merged with the other polyhedra, its result does not.
Indeed, the fact that two polyhedra have a non-empty intersection will stay true even
if one or both of them are merged with other polyhedra.
This result, as depicted in Section~\ref{approach:duplex}, is an over-approximation
of disjoint areas of the gray state space.

In practice, both the ECH based hullification and the ICH calculation heuristics
run in parallel, and the sets of potential invariants they output are merged
before being sent to the $k$-induction.
This allows us to combine the precision of ECHs with the over-approximation effect
of ICHs.
A high level view of our approach is available on Figure~\ref{fig:global}.
The next section will present two examples taken from a functional chain as
presented in Section~\ref{systems} each illustrating the ideas introduced in this
section: a reconfiguration logic system and a voting logic system.

\begin{figure}
  \begin{centering}
  \scriptsize
  \begin{tikzpicture}[scale=.8,auto,node distance=2cm,>=latex]
    \draw[draw=black,fill=blue!20] (0,2) -- (0,6) -- (10,6) -- (10,2) -- cycle;
    \node (HQ) at (0.7,5.7) {HullQe};
    \node[anchor=west] (Key) at (0.65,1)
    {\begin{tabular}{c c r}
    $F_i(s,s')=$ & $PO(s)\land invs(s)\land T(s,s')\land invs(s')\land\neg PO(s')$ & if $i=1$\\
                 & $PO(s)\land invs(s)\land T(s,s')\land invs(s')\land G_{i-1}(s')$ & if $i>1$\\
    \end{tabular}};
    \draw[draw=black] (1.45,6) -- (1.45,5.4) --(0,5.4);
    \node[draw=black,fill=blue!10] (BQE) at (1,4)  {HullQe};
    \node[draw=black,fill=blue!10] (QE)  at (4,4)  {QE};
    \node[draw=black,fill=blue!10] (ICH) at (5,5.5)  {ICH};
    \node[draw=black,fill=blue!10] (ECH) at (5,2.5)  {ECH};
    \node[draw=black,fill=blue!10,text width=40pt,text centered] (Fla) at (8,4)  {Atom extraction and negation};
    \node[draw=black,fill=blue!10,text width=70pt,text centered] (Kin) at (13,4) {$k$-induction};
    \node[draw=black,fill=blue!10] (Dis) at (13,5.5) {Discarded};
    \node[draw=black,fill=blue!10] (PO)  at (13,2.5) {Contains PO};
    \node[draw=black,fill=blue!10] (Do)  at (15,2.5) {Done};

    \draw[->, >=latex] (QE) -- node[anchor=south] {$G_i$} (BQE);
    \draw[->, >=latex] (BQE.north) ..controls +(up:10mm) and +(up:10mm).. node {$F_i(s,s')$} (QE.north);
    \draw[->, >=latex] (QE) -- node[anchor=west,pos=.1] {$\bigvee G_i$} (ICH);
    \draw[->, >=latex] (QE) -- node[anchor=east] {$\bigvee G_i$} (ECH);
    \draw[->, >=latex] (ICH) -- node {inexact hulls} (Fla);
    \draw[->, >=latex] (ECH) -- node[anchor=west,pos=.01] {exact hulls} (Fla);
    \draw[->, >=latex] (Fla) -- node {$atoms$} (Kin);
    \draw[->, >=latex] (Kin.0) ..controls +(1,0) and +(1,1).. node[anchor=west,text centered] {$U$ (retried on next iteration)} (Kin.45);
    \draw[->, >=latex] (Kin) -- node {$F$} (Dis);
    \draw[->, >=latex] (Kin) -- node[name=v] {$V$} (PO);
    \draw[->, >=latex] (PO) -- (Do);
    \draw[->, >=latex] (v) ..controls +(left:3mm) and +(down:33mm).. node[pos=0.97,anchor=west] {invariants} (BQE.south);
  \end{tikzpicture}  \caption{High level sequential description}
  \label{fig:global}
  \end{centering}
\end{figure}

%% file: Applications.tex
\label{applications}

In this section we discuss the results of the proposed approach on two
real world examples: a reconfiguration logic and the triplex voter of
Rockwell Collins.

\subsection{Reconfiguration Logic}
\label{applications:reconf}

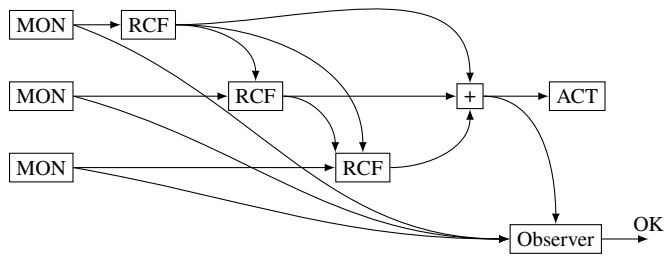
\begin{figure}[t]
\begin{center}
\scriptsize
\begin{tikzpicture}[scale=0.95]
\node[draw] (W1) at (-3.0, 2.0) {MON};
\node[draw] (W2) at (-3.0, 1.0) {MON};
\node[draw] (W3) at (-3.0, 0.0) {MON};
\node[draw] (RCF1) at (-1.5, 2.0) {RCF};
\node[draw] (RCF2) at ( 0.0, 1.0) {RCF};
\node[draw] (RCF3) at ( 1.5, 0.0) {RCF};
\node[draw] (PLUS) at ( 3.0, 1.0) {+};
\node[draw] (ACT) at  ( 4.5, 1.0) {ACT};
\node[draw] (OBS) at  ( 4.2, -1.0) {Observer};

\draw[->, >=latex] (W1) -- (RCF1);
\draw[->, >=latex] (W2) -- (RCF2);
\draw[->, >=latex] (W3) -- (RCF3);

\draw[->, >=latex] (RCF1) to[out=0, in=90] (PLUS.north);
\draw[->, >=latex] (RCF1) to[out=0, in=90] (RCF2.north);
\draw[->, >=latex] (RCF1) to[out=0, in=90] (RCF3.north);

\draw[->, >=latex] (RCF2) to[out=0, in=180] (PLUS.west);
\draw[->, >=latex] (RCF2) to[out=0, in=90] (RCF3.north west);
\draw[->, >=latex] (RCF3) to[out=0, in=-90] (PLUS.south);

\draw[->, >=latex] (PLUS) -- (ACT);

\draw[->, >=latex] (PLUS.east) to[out=0, in=90] (OBS.north);
\draw[->, >=latex] (W1.east) to[out=-20, in=180] (OBS.west);
\draw[->, >=latex] (W2.east) to[out=-15, in=180] (OBS.west);
\draw[->, >=latex] (W3.east) to[out=-10, in=180] (OBS.west);
\draw[->, >=latex] (OBS.east) -- (5.5,-1.0)node[above]{OK};
\end{tikzpicture}
\end{center}
\normalsize
\caption{Reconfiguration subsystem with observer.}
\label{fig:reconf-observer}
\vspace{-15pt}
\end{figure}

Distributed reconfiguration logic as presented in
Section~\ref{systems} would be best described as a distributed
priority mechanism. In each redundant channel, the reconfiguration
logic comes last and monitors the warning flags raised by the
monitoring logic implemented earlier in the data flow. Integer timers and latches are used to confirm warnings over a number of
consecutive time steps and trigger a reconfiguration. The duration of
the various confirmations can vary from a few steps to hundreds or
thousands of steps and are tuned by system designers to be not overly
sensitive to transient perturbations, which would unnecessarily trigger
reconfigurations of otherwise healthy channels, while being fast
enough to ensure safety. Assuming at most two
sensors, network or CPU faults, the following generic property is expected
to hold for the reconfiguration mechanism: ``No unhealthy channel shall be in
control for more than $N$ steps''. This property can be decomposed and
instantiated per channel. However, a property such as ``No more than one channel shall be in command at any time'', or ``The actuator must never stay idle for
more than $m_4$ steps'' are more challenging because they cover all
three channels simultaneously and drag many state
variables in their cone of influence. For instance, the formal verification of the second
property is done by
assembling a model of the distributed system and by using the
synchronous observer technique as shown in Figure~\ref{fig:reconf-observer}. The observer uses a timer and is coded
so that its output becomes true as soon as the absence of control of
the actuator has been confirmed for the requested amount of $m_4$
consecutive steps. The proof objective on the system/observer composition is
to show that the output of this observer can never be true.

The timer logic found in this system is similar to that of the toy example developed in
Section~\ref{approach:counter}, and instantiated several times, indeed a channel becoming corrupt triggers several timers with different
bounds, running into each other or in parallel.
Let us now see how  hullification performs on this system.
The first preimage does not contain enough
information, since hullification generates no potential lemma which either strengthens
the PO or is $k$-inductive by itself.
The union of the first and second preimages however allows hullification to generate
about $200$  potential invariants.
Once they are negated, $k$-induction invalidates most of them
and indicates the PO was found ($1$-)inductive conjoined with about $50$ lemmas after about $30$ seconds of computation.

After the minimization phase described in Section~\ref{approach}, it turns out
that only three lemmas are required.
If we call $timer_i$ the integer variable used to count the time
channel $i$ is not in command for $1\leq i\leq 3$, and $timer_o$
the timer used by the observer, the lemmas are:
$\neg(timer_o - timer_i\;\geq m_4 - m_i -1)$
where $1\leq i\leq 3$.
These lemmas are found no matter the values of the $m_i$
for $1\leq i\leq 4$.
We insist on the interest of hullification here.
Merging polyhedra in some single arbitrary order is too coarse and the resulting
hull cannot strengthen the PO, whereas the thorough exploration generates useful lemmas.

The reconfiguration logic was also analyzed using NBac, Scade Design
Verifier and Tinelli's Kind.  NBac did not succeed in proving the
property after 1 hour of computation. Both the Scade Design Verifier
and Kind kept on incrementing the induction depth without finding a proof after 30 minutes of run time.

The invariant generation of Kind was also run on this system, and
yielded a number of small theorems, but obviously not property
directed and unfortunately not sufficient to strengthen the PO and
prove it.

In conclusion, the proposed combination of backward analysis, hullification and $k$-induction allows us to complete a
proof in a few seconds on a widely used avionics design pattern, where other
state of the art tools fail. In addition, we see two very interesting points worth
highlighting about hullification:
\begin{inparaenum}[\itshape (i)]
  \item The PO is made ($1$-)inductive, implying the proof can
        easily and quickly be re-run and checked by any existing induction tool;
  \item the time needed to complete the proof does not depend
        on the numerical values of the system --about thirty
        seconds on a decent machine in practice for this
        system\footnote{Using our prototype implementation in Scala.}.
\end{inparaenum}
This is very important for critical embedded systems manufacturers
as point {\itshape (i)} means that the proofs are trustworthy, both for the industrials
themselves and the certification organisms.
On the other hand, point {\itshape (ii)} implies that strengthening lemmas can be very quickly generated
for similar design patterns with altered numerical values, easing the
integration of formal verification in the development process.
Indeed, it avoids the need for an expert to manually transpose the lemmas on the
new system, as can be the case for complicated and resource/time consuming
proofs.

\subsection{The Triplex Voter}

\label{applications:voter}

Let us now turn to the Rockwell Collins triplex sensor voter, an
industrial example of voting logic as introduced in
Section~\ref{systems}, implementing
redundancy management for three sensor input values. 
This voter does not compute an average value, but uses the $middleValue(x,y,z)$ function, 
which returns the input value, bounded by the minimum and the maximum input values
({\em i.e.} $z$ if $y < z < x$).
Other voter algorithms which use a (possibly weighted) average value are more sensitive 
to one of the input values being out of the normal bounds.
The values considered for voting are {\em equalized} by subtracting
equalization values from the inputs.
The following recursive equations describe the behaviour of the voter with $X \in \{A, B, C\}$:
\[
\begin{array}{rcl}
EqualizationX_0 & = & 0.0\\
EqualizedX_t & = & InputX_t - EqualizationX_t\\
EqualizationX_{t+1}	& = & 0.9 * EqualizationX_t + \\
& & \hspace{-12ex}0.05 * (InputX_t + ((EqualizationX_t - VoterOutput_t) - Centering_t))\\
Centering_t & = & middleValue(EqualizationA_t, EqualizationB_t,\\
& &\hspace{14.5ex} EqualizationC_t)\\
VoterOutput_t & = & middleValue(EqualizedA_t, EqualizedB_t, EqualizedC_t)\\
\end{array}
\]
The role of the equalization values is to compensate offset errors of 
the sensors, assuming that the middle value gives the most accurate 
measurement. 

We are interested in proving Bounded-Input Bounded-Output (BIBO) stability 
of the voter, which is a fundamental requirement for 
filtering and signal processing systems, ensuring that the system output cannot 
grow indefinitely as long as the system input stays within a certain range. In 
general, it is necessary to identify and prove auxiliary system invariants in 
order to prove BIBO stability.

So, we want to prove the stability of the system, {\em i.e.} we want to
prove that the voter output is bounded as long as the input values differ 
by at most the maximal authorized deviation $MaxDev$ from the true value 
of the measured physical quantity represented by the variable $TrueValue$.
In our analysis, we fixed the maximal sensor deviation to $0.2$, a value 
that domain experts gave us as typical value in practical applications.
It is staightforward to prove that the system is stable if the equalization 
values are bounded. 

When applied to Rockwell Collins \textbf{tri}plex sensor voter, our prototype implementation
manages to prove the PO in less than $10$ seconds by discovering that
$-0.9\le \sum^3_{i=1} Equalization_i\le 0.9$ is a strengthening lemma, using ICH calculation.
Again, the time taken to complete the proof does not depend on the
system numerical constants, and the strengthened PO is ($1$-)inductive.
We insist on the importance of these characteristics for both industrials and
certification organisms: the proof is trustworthy and can be redone easily for
similar, slightly altered designs.

The stability of the system without fault detection nor reset was already 
proven in~\cite{dierkes11}, but the necessary lemmas had to be found by hand
after the Scade Design Verifier, Kind as well as Astrée (which was run
on C-Code generated from the Lustre source) failed at automatically
verifying the BIBO property.

%% file: Framework.tex
\label{framework}
Our actor oriented collabo\-rative
verification framework~\cite{saeChampion} called {\em Stuff}
is composed of several elements: the $k$-induction engine, the
abstract interpreter, the backward analysis, ICH calculation and ECH hullification.
They can all evolve in parallel and communicate.

Stuff is written in Scala except for the abstract interpreter, written in OCaml.
We implemented the backward analysis and the two heuristics presented in this paper
using the QE algorithm from~\cite{DBLP:conf/lpar/Monniaux08} modified to handle integers or reals whith booleans.
The underlying projections of~\cite{DBLP:conf/lpar/Monniaux08}
are performed by the Parma Polyhedra Library~\cite{BagnaraHZ08SCP}, also used for convex hull
computation.
Stuff can use any SMT lib 2.0~\cite{BarST-SMT-10} compliant solver thanks to the
Assumptio\footnote{\href{https://cavale.enseeiht.fr/redmine/projects/assumptio}{https://cavale.enseeiht.fr/redmine/projects/assumptio}}
actor oriented SMT solver wrapper.
In practice, the backward analysis and the heuristics use Microsoft Research Z3~\cite{DBLP:conf/tacas/MouraB08}
and MathSat~$5$~\cite{mathsat5} by the University of Trento.

A run of the framework in the default configuration begins by a preprocessing phase using
abstract interpretation with intervals as abstract domains in order to
infer bounds on the state variables. This provides an over-approximation
of the reachable state space which once verified by $k$-induction
is propagated to all the other elements of the framework. The rest of the analysis
follows the approach discussed in Section~\ref{approach} with the backward analysis
feeding preimages to both ECH hullification and ICH calculation.
They in turn feed potential invariants to the $k$-induction engine which
detects real invariants and check if they strengthen the property as described
in Section~\ref{approach}.
In this setting, even if our approach does not consider
the initial states, it benefits from the over-approximation
of the AI preprocessing phase, which takes into account the
initial states but not the PO. The AI results also enhance the quality
of the output and the overall performance of the incremental
$k$-induction engine.

%% file: Conclusion.tex
\label{conclusion}

In this paper, the authors presented two automatic and property directed lemma
generation heuristics, which operate on preimages of the negation of the proof objective obtained by a backward exploration, itself powered by quantifier elimination.

The first heuristic originality lies in the thorough
exploration of a set of possible convex partitionings of the gray
state space by exact convex hull calculations.
This exploration, called {\em hullification}, is performed
incrementally, as soon as new preimages containing new information about the
gray state space, are computed by the backward analysis.
As illustrated on the reconfiguration logic example, the blowup inherent to the
exploration of the partitionings is avoided thanks to the optimizations
discussed in this paper and far outperforms other available tools.

The second heuristic over-approximates disjoint areas of the gray state space
by accepting inexact hulls when the candidate polyhedra intersect.
It performs very well in the Rockwell Collins Triplex Sensor Voter experiments,
allowing to conclude a proof none of the other state of the art tools could conclude.

These results, obtained with the prototype
implementation of the proposed method, are of interest in our application field. Indeed, they
allow to discover strengthening lemmas, in reasonable time, for
essential safety properties of widely used fault tolerance design patterns at
model level, a task which has proved difficult to achieve using other
techniques such as AI or $k$-induction with manual analysis of failed proofs.

Future work include further reflexion on systems mixing integers and
reals and on heuristics using preimages from the backward analysis.
Also, the authors think that when hullification cannot find strengthening lemmas,
it can still provide interesting starting points for template based
techniques and experiments have been started in this direction.
Outside of the proposed approach,
the authors believe in a multi method approach and will continue to experiment
in this direction: work on an
implementation of PDR~\cite{DBLP:conf/vmcai/Bradley11,een2011}
adapted to numerical systems is in progress. It was
observed that PDR is able to discover range lemmas similar to those found
using interval based AI, while being able to conclude inductive proofs,
and the cooperation of hullification and PDR is being studied.
The long term goal is to refine and bridge the
verification techniques developed for precise parts of the
functional chains (voting, reconfiguration logic and numerical
stability for control laws) to obtain a methodology and tool support suitable
for end-to-end verification of avionics software at model level.